# Trimming and ultra-wide bandwidth expansion of silicon frequency comb spectra with self-adaptive boundary waveguides

*Jianhao Zhang, \*,†,§ † Vincent Pelgrin,† Carlos Alonso-Ramos,† Laurent Vivien,† Sailing He,,§*

*and Eric Cassan,\*,*

† Centre for Nanoscience and Nanotechnology (C2N), CNRS, University Paris-Sud,

University Paris-Saclay, 91405 Orsay cedex, France

§Centre for Optical and Electromagnetic Research, Zijingang Campus, Zhejiang University,

Hangzhou 310058, China

\* Contribute equally to the work.

eric.cassan@u-psud.fr

ABSTRACT: Dispersion engineering is among the most important steps towards a promising optical frequency comb. We propose a new and general approach to trim frequency combs using a self-adaptive boundary of the optical mode at different wavelengths in a sub-wavelength structured waveguide. The feasibility of ultra-wide bandwidth dispersion engineering comes from the fact that light at different wavelengths automatically self-adapts to slightly different effective spatial spans determined by the effective indices of the mode. Using this self-adaptive variation on the confinement, we open up the window of low-anomalous dispersion in a large wavelength range, and theoretically demonstrate frequency combs with improved bandwidths with respect to the state-of-art in several different waveguide configurations considered, for a matter of illustration, in the silicon photonic platform. This strategy opens up a new design space for trimming the spectrum of frequency combs using high-index-contrast platforms and provides benefit to various versatile nonlinear applications in which the manipulation of energy spacing and phase matching are pivotal.

**Introduction**

Nonlinear processes have raised a great interest due to their unique capabilities for on-chip light generation and spectral/time domains manipulation, with an immense potential for the implementation of light sources in silicon photonics [1-4] based on optical parametric amplification and supercontinuum and frequency comb effects based on $3^{rd}$ nonlinear Kerr effect [5-9]. As an emerging research topic that has strong potential in achieving on-chip broadband light sources, frequency comb generation based on nonlinear Kerr effect (i.e. cascaded four-wave mixing) has recently drawn significant attention [10-19]. Consisting in discrete and equally spaced frequency lines spectra, Kerr frequency combs strongly request the precise control of the dispersion and nonlinearity, gain and loss on optical waveguides [20], especially when temporal patterns with few solitons are expected.

Controlling the dispersion of optical waveguides is a very important preliminary step for the exploitation of Kerr frequency combs, for which great effort was devoted to compensate both the intrinsic material and the nonlinearity-induced dispersions in order to satisfy the energy and momentum conservation conditions. To balance the nonlinearities-induced phase mismatch, overall anomalous dispersion is generally expected [10, 11, 20], which can be supported directly from the materials (e.g. silica at telecom wavelengths) or induced by the waveguide dispersion with well-designed waveguide cross-sections. As a result, toroidal-shape cavities using silica [10] or $MgF_2$ [12, 13] are frequently used for frequency comb generation due to the high $Q$ factor of up to few millions and low dispersion of these structures. Another classical material platform used for frequency comb demonstrations is silicon nitride (SiN). Due to the high quality factor of ~ million and a much higher nonlinear index (SiN: $n_2 \approx 2.4 \times 10^{-19} m^2/W$, $MgF_2$: $n_2 \approx 1.5 \times 10^{-20} m^2/W$), high-performance frequency combs could be achieved on chip [15-18] and even integrated with commercial III-V integrated lasers [19]. Though silicon nitride presents normal material dispersion at around ~$1.5\mu m$ wavelength, its high enough index contrast with $SiO_2$ allows it to generate small anomalous dispersion to compensate nonlinear-induced phase mismatch.

However, achieving broadband phase-matching in simple-to-fabricate and fabrication-tolerant silicon waveguides remains an open question, since the ultra-high index contrast between Si and $SiO_2$ leads to a rapid evolution of the dispersion curves [21, 22]. Broadband phase-matching has been shown based on optimization of high order diffraction terms, e.g. 4th order dispersive waveguides [15, 23], or by implementing rib geometries [24]. Yet, the proposed solutions require complex fabrication processes, with deposition of different materials, or tight control of rib and slab thicknesses. A flexible method to flatten silicon waveguide dispersion for frequency comb generation is thus highly expected. A strong interest is thus to explore how the dispersion shape can influence the comb bandwidth as well as other properties of the comb spectrum, especially in situations when a target wavelength operation is provided in advance.

We propose here a universal method to trim the dispersion of strong index contrast waveguides using a self-adaptive boundary condition (defined hereafter) in order to significantly extend the bandwidth of comb spectra. The needed self-adaptive boundary was originally introduced in our another article [25] for multimode optical waveguides, where equivalent potential wells [26-28] in the waveguide transverse direction were achieved by specially-designed lateral index profiles, providing room for light modes to automatically self-adapt to slightly different effective spatial spans. We explore here the properties of such guides in single mode operation and demonstrate that they have very interesting properties for dispersion engineering in a large wavelength range, thus providing new room for trimming the spectra of frequency combs using high-index-contrast platforms. The feasibility of spectrum engineering further enhances optical frequency comb as a strong candidate for novel on-chip silicon light sources [29], and provides a strong basis for many applications in on-chip spectroscopy or metrology [30], as well as novel research axes such as time-space-frequency mapping [31].

**Comb bandwidth limited by the dispersion of silicon strip waveguide**

The generation of frequency combs using ring microresonators is traditionally done from the configuration shown in figure 1, for which the overall process of comb generation [12, 32-47] was systematically discussed in past ten years and briefly recalled here. The propagation of short light pulses in waveguides for nonlinear optics has been extensively described in the literature, starting with fibers [32] and then, for more than 10 years, in the higher contrast index waveguides of integrated photonics, particularly in silicon photonics [20]. In summary, the propagation of pulses with simultaneous formation of an optical frequency comb in a waveguide with effective third order non-linearity can be described by non-linear Schrödinger equation (NLSE) [36-41, 43 ,47]:

$$\frac{\partial}{\partial z}A(z,\tau) = \left[i\sum_{k=2}^{n}\frac{\beta_k}{k!}(i\frac{\partial^k}{\partial \tau^k}) + i\gamma|A(z,\tau)|^2 - \frac{\beta_{2PA}}{2A_{eff}}|A(z,\tau)|^2 - \frac{\beta_{3PA}}{3A_{eff}^2}|A(z,\tau)|^4 - \frac{\alpha}{2}\right]A(z,\tau)$$

(1)

constrained by the boundary condition from ring in/out coupling from a bus waveguide:

$$A(0,\tau)_{m+1} = \sqrt{1-\kappa}A(L,\tau)_m e^{i\Delta\varphi_0} + \sqrt{\kappa}A_{in}(\tau) \quad (2)$$

, where $A(z,\tau)$ is the pulse amplitude described by the circumferential position $z$ and the time variable $\tau$ corresponding to a relative time frame. $\beta_k$, $\gamma$ and $\alpha$ are the group velocity dispersion, Kerr nonlinear parameter and waveguide loss, respectively while $\beta_{2PA}$, $\beta_{3PA}$ and $A_{eff}$ are the two-photon, three-photon absorption coefficient and effective mode area. $A_{in}$ is the driven amplitude while $\Delta\varphi_0$ is the phase detuning satifying $\Delta\varphi_0 = L\omega_0 n_g/c = 2\pi n - \varphi_0$ where $\omega_0$ and $\omega_n$ are the pump frequency and $\omega_n$ the closest $n$-order resonant frequency of ring, respectively.

$\varphi_0$ is the phase of the driven signal accumulated in the cavity. For convenience, $n$ is usually set with zero order by which $\Delta\varphi_0 = -\delta_0$. Driven wavelength red shifted or blue shifted are described by $\delta_0 > 0$ and $\delta_0 < 0$, respectively. It is further developed into a two-time scale approach, in the form of the Lugiato–Lefever equation (LLE) [35-38], where a fast time variable describes the pulse field profile in a reference frame moving at the group velocity and spanning in the roundtrip time range, while another one measures the evolution averaged field of the round-trip. i.e:

$$\tau_R \frac{\partial}{\partial t} A(t,\tau) = \sqrt{\kappa} A_{in}(\tau) - \left(\frac{\alpha L + \kappa}{2} + i\delta_0\right) A(t,\tau) + i \sum_{k=2}^{n} L \frac{\beta_k}{k!} \left(i \frac{\partial^k}{\partial \tau^k}\right) A(t,\tau) + i\gamma L |A(t,\tau)|^2 A(t,\tau) - \frac{\beta_{2PA} L}{2 A_{eff}} |A(z,\tau)|^2 - \frac{\beta_{3PA} L}{3 A_{eff}^2} |A(z,\tau)|^4$$

(3)

This approach widely is referred as the ***mean-field Lugiato–Lefever equation***, which can be solved by doing integration or searching the possible roots with the Newton-Rhapson method.

A typical spectrum of an on-chip SiN-based frequency comb [11] is shown in figure 1 (b) in which anomalous dispersion for broadband comb spectrum should be controlled by properly choosing the waveguide cross-section. As for us, our approach started from there, and we developed a code for solving the LLE equation, whose stability and accuracy we first validated in comparison with previous results from the literature [11]. With the dispersion and other parameters (including the micro-ring $Q$ factor, FSR, and the pump power level) from [11], we successfully reproduced the main results, achieving different single-soliton regimes as the single-soliton on shown in figure 1 (c) Then, we employed it to specifically explore the behavior of silicon frequency combs and introduce our improvements for addressing several drawbacks of silicon frequency combs related to previous limitations in term of phase-matching bandwidth.

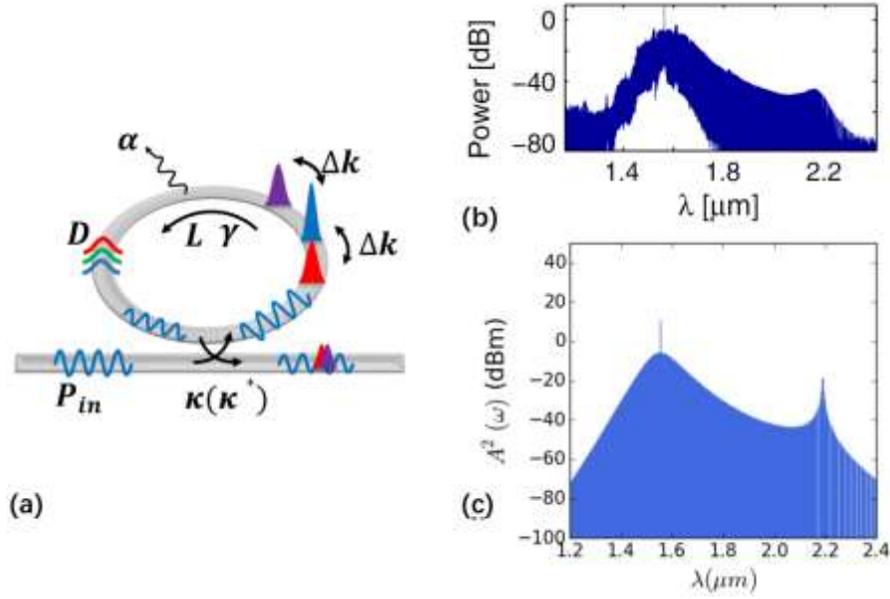

**Fig. 1.** (a) Schematic of the nonlinear ring-based frequency comb system. (b) experimental comb spectrum generated in a silicon nitride ring resonator, from [11]. (c) A recovered dispersion and (d) spectrum of a single soliton state of frequency comb based on the same parameters.

To explore silicon frequency comb, we first calculate the dispersion $D$ parameter of SOI waveguides with different dimensions, as shown in figure 2. Once the pump wavelength is chosen, the anomalous dispersion ($D > 0$) can be created by increasing the waveguide height and later be flattened with larger and larger widths. In the meanwhile, the peak position of the parabolic shape dispersion curve is then red shifted. Absolute values and peak positions are controlled by the height and width of the waveguides, and it turns that dispersion curves rapidly change from negative to positive regions due to the high index contrast of SOI waveguides. To obtain a small anomalous dispersion region needed to phase matching purpose, waveguide dimensions need to be very carefully controlled or external materials should be used for dispersion compensation, but these two approaches are neither easy to perform nor fully satisfying [14].

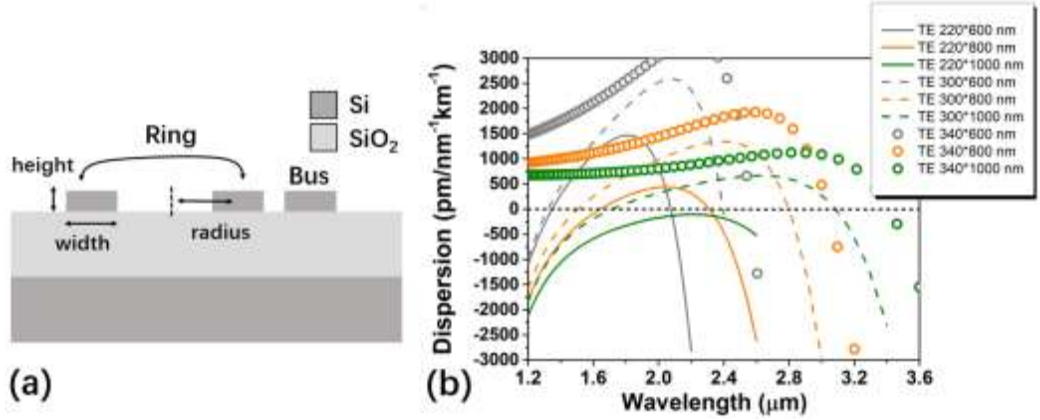

**Fig. 2.** (a) Cross-section of the silicon ring used for comb modeling. (b) Dispersion curves of silicon-on-insulator waveguide with different cross-section.

For a waveguide with 220nm height and 800nm width (in order to have low anomalous dispersion around 2.2µm wavelengths), the waveguide dispersion curve displays a peak value of chromatic dispersion around $D = 350\ pm/(nm * k)$, as figure 3 (a). The corresponding group velocity dispersion is shown in figure 3 (b). An underestimated nonlinear gain coefficient of $\gamma = 30(m \cdot W)^{-1}$ is considered here for interrogating the other parameters, especially the pump power. A coupling coefficient of 0.01 (working in under coupling) and 150mW pump power (in bus waveguide) are chosen, respectively to overwhelm the losses. A free spectrum range of 0.23 THz corresponding to a ring radius of $50\mu m$ is chosen to build a moving frame with roundtrip time of $4.35\ ps$. Using the mean-field LLE model described above, we achieved a single-soliton silicon frequency comb with intracavity pulse shape shown in figure 3 (c). Compared to those silicon nitride configuration [11], the intracavity energy inside silicon is much lower due to the much higher gain factor (from 1 to $30(m \cdot W)^{-1}$), which means that much less power is now required to overwhelm propagation losses. This is also a strong point and original interest of using the silicon platform to achieve frequency combs. However, it can be seen in figure 3 (d) that, due to the much higher waveguide dispersion, the bandwidth of the silicon frequency comb is much worse than the one of silicon nitride.

To summarize the current point of our discussion, it appears that by moving from $Si_3N_4$ to SOI waveguides, an advantage (lower pump power) and a drawback (reduced frequency comb spectral width) result simultaneously. Consequently, addressing the problem of the spectral width of µ-combs is an important point, naturally aiming to increase it by using waveguides with a particular chromatic dispersion profile.

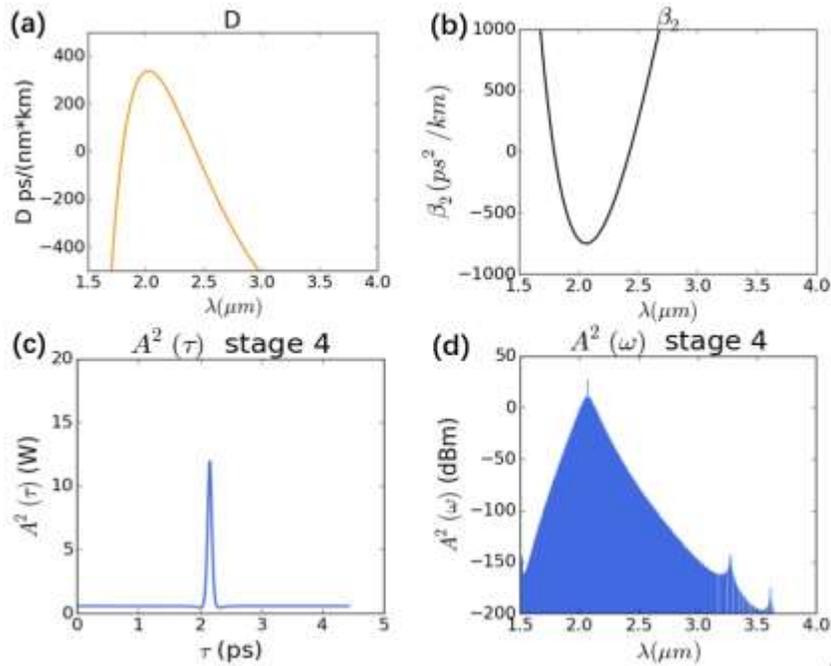

**Fig. 3.** (a) Dispersion and (b) Group velocity dispersion of the SOI waveguide that shown in Figure 2. (c) Final soliton stage and (d) Corresponding soliton frequency comb.

To understand the bandwidth issue, we come back here to simulation results. In figure 4 (a), the slightly red tuning gives birth to a situation where four-wave mixing (FWM) parametric gain overwhelms the overall loss level. The first primary frequencies generated by degenerate FWM, as in figure 4 (a), are located very closely to the pump line. These primary lines are actually the pre-representation of the comb bandwidth since the following cascaded FWM and mode locking processes all originate from there. A snapshot of the separated sub combs that result from sufficient cascaded FWM mixing processes is shown in figure 4 (b). Due to the strong dispersion $D$, FWMs are limited to small range according to phase matching, which can be described as:

$$\Delta k = 2\gamma P_{in} - D\frac{\lambda^2}{2\pi c} \cdot (nFSR_\omega)^2 = 2\gamma P_{in} - D2\pi c \cdot (\frac{n\lambda}{n_g L})^2 \qquad (4)$$

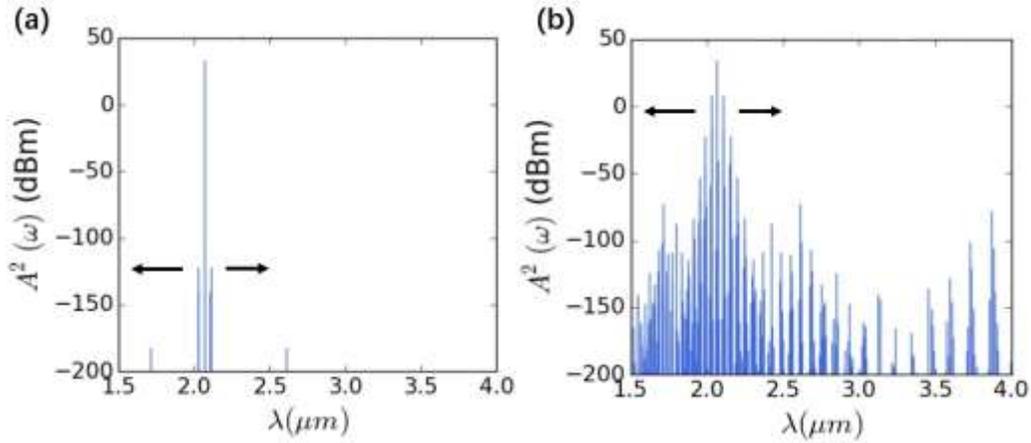

**Fig. 4.** (a) Snapshot of the generation of primary frequency lines due to the degenerate four-wave mixing. (b) A later snapshot where multi frequency lines by are generated by cascaded four wave mixing.

,for degenerate FWM with three waves spaced by one FSR ($n = 1$), the pump power for phase matching (following the same configuration, $\lambda = 2.1\mu m$, ring $L = 314\mu m$ and $\gamma = 30m^{-1}W^{-1}$) is 27.8mW. In another words, for a pump power that we use (150mW), the phase matching occurs at the frequency lines of 2.8 FSR from pump, which is coincident to what we observe in Figure 3 (a). So very straightforwardly, if we want to move the primary frequency lines to a further place for extending the bandwidth (black arrows in Figure 4), a much stronger power, which responds to the square-order growth of FSR number $n$, is needed to compensate the dispersion. This high power is always wasting since it is far overwheled by optical losses and not accesible. On the contrary, it is very significant and interesting to reduce and flatten the dispersion $D$ globally while keeping the other parameters fixed ($\gamma$, $P_{in}$ etc.).

**Silicon frequency comb with engineered dispersion in self-adaptive boundary waveguide**

In a step-index waveguide like in figure 5 (a), all the wavelengths are feeling the same physical boundary. When wavelength is continuously raised to where a big part of the energy is pushed out to the unlimited cladding bulk material, the normal dispersion produced by the left part starts to dominate the total. The dispersion curve of SOI waveguide without material dispersion is presented in Figure 5 (b). The waveguide could be understood like a road in which the volume vehicles is the wavelength. With different vehicles on the same road, a driver makes difference on stepping on the gas (dispersion) unless he/she feels the same compatibility (confinement) on the road. If one wants to travel with same speed, the very simple idea is to diversify the "road" for different "vehicles", which is exactly what the Self-Adaptive Boundary condition (SAB) is made for.

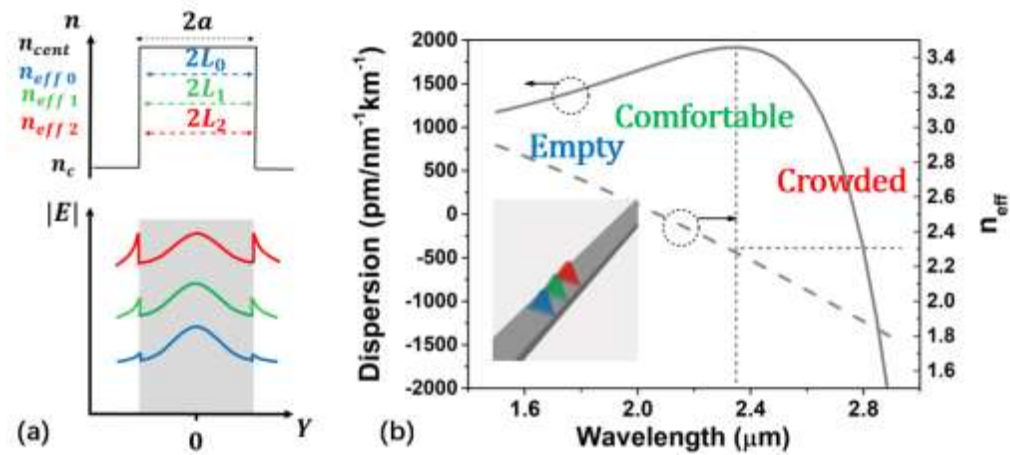

**Figure 5.** (a) Schematic of a step-index waveguide. (b) The chromatic dispersion parameter (D) of a SOI waveguide with a width and height of 750nm and 300nm respectively. The material dispersion is not considered here ($n = 3.48$).

With this objective in mind, we have endeavored to explore the use of SAB waveguides [25], whose dispersion properties intended to be adapted to the problem of µ-combs. Figure 6 shows two types of waveguides with Self-Adaption Boundary condition $n_{eff} > n_b$, i.e. continuously graded-index waveguide (Figure 6 a) and discretely graded-index waveguides (Figure 6 c), in single-mode operation.

The structure of the bi-level graded-index waveguide is shown in the inset of Figure 6 (a). The period thickness and width are chosen at 150nm, 340nm and 800nm, with a filling factor linearly tapered from center to edge. Dispersion of a strip waveguide and SAB waveguide with $n_b \approx 1.5$ and $n_b \approx 2.8$ is plotted in figure 6 (b) for comparison. The SAB allow us to trim and expand the dispersion of the "long" wavelength where $n_{eff} < n_b$. This is because the wave is confined by the index contrast of $n_b/n_c$ and phase integral strongly depends on the index $n_b$. In contrast, as the short wavelength ($n_{eff} > n_b$) is confined by the effective width of the waveguide where electric field smoothly evolves, the waveguide dispersion does not vary a lot. In brief, the whole frequency range is separated into two parts with diverse responses and this allows us to globally reconfigure the dispersion by flattening or steepening the dispersion accordingly. Using the step-index graded-index waveguides shown in figure 6 (c), one can even further extend the anomalous dispersion range by locally squeezing a small anomalous dispersion. The bi-level SAB is used for confining separately the "blue" and "red" wavelength. This idea [25] can practically be translated into a subwavelength structured waveguide [48-51] as shown in Figure 6 (d).

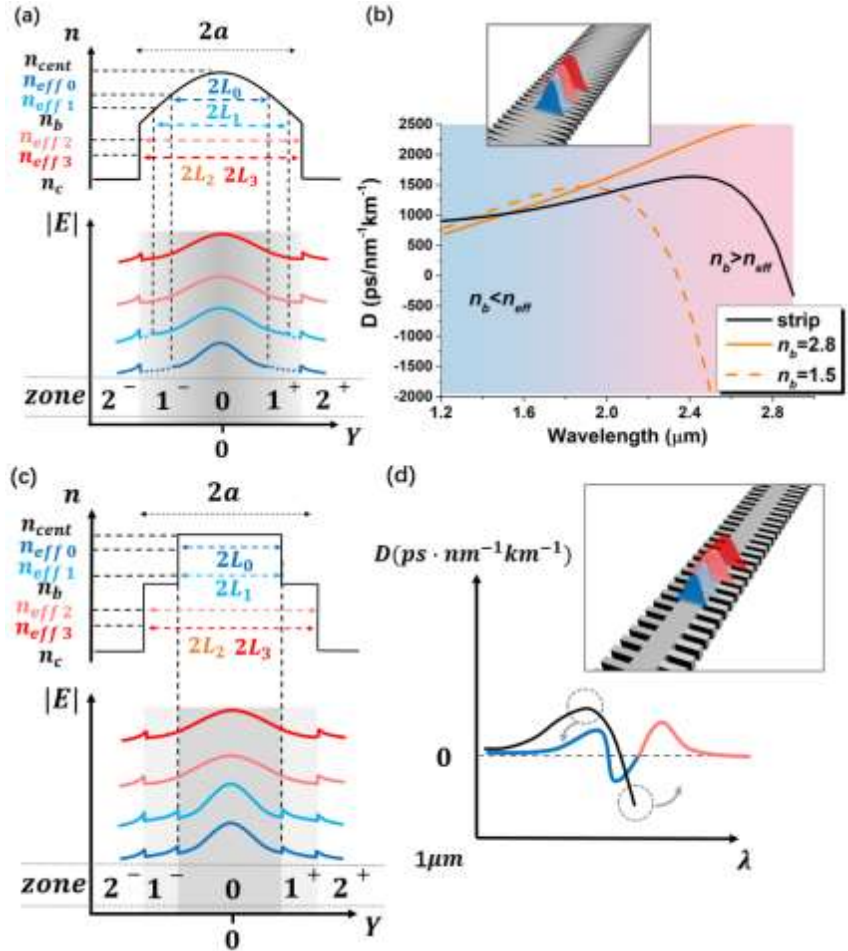

**Fig. 6.** (a) Schematic of a graded-index waveguide with SAB operating in single-mode. (b) Chromatic dispersion parameter D. Width and height of the waveguide are 750nm and 340nm. (c) Schematic of a bi-level graded-index waveguide with SAB operating in single-mode. (d) Sketch of the chromatic dispersion parameter D. Material dispersion is not considered here ($n = 3.48$).

A commercial 340nm SOI platform was chosen to provide sufficient anomalous dispersion in up to $2.2\mu m$ range, where two-photon absorption can be nearly eliminated. Since the SAB condition can reduce the index contrast and dramatically reduce the anomalous dispersion for wavelengths not confined by the outer boundary, the region between two anomalous peaks are seperated by a normal region using a bi-level SAB waveguide, as shown in Figure 7(a). In this case, a more appropriate index contrast with filling factor varying from 1 to 0.66 is achieved (dimension of the waveguide is clarified in the inset). The whole wavelength range from $2.25\mu m$ to 3.25$\mu m$ are now totally flattened. Compared to those classical strip/rib waveguides where the dispersion displays normally a parabolic shape, there is more than one peak existing in the center which totally gives us a new freedom for the dispersion engineering. In addition to the bi-level case, tri-level case could be further use to locally engineer the dispersion, as shown in Figure 7 (a). The small ripple can be eliminated by adjusting the length of each section once the filling factor is determined, as illustrated in figure 7 (b). In a wide wavelength range, the anomalous dispersion can thus be controlled and limited to very low values.

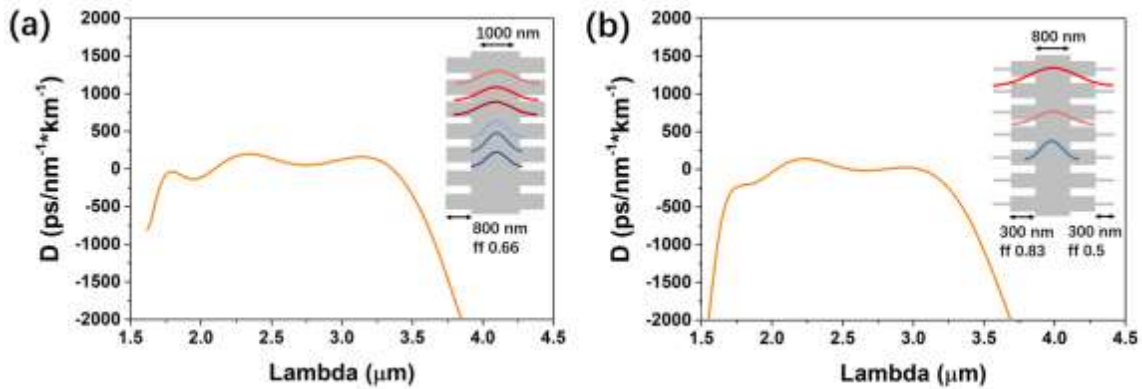

**Fig. 7.** Dispersion of infinite-level (a), bi-level (b) and three tri-level (c) (d) subwavelength structured waveguides. The thickness of silicon and the period are both set at 340nm and 240nm. Other parameters are respectively shown in the insets.

Frequency comb generation was then considered with the flattened dispersion waveguides shown in figure 7. Since periodic structures are introduced in this case, extra losses are needed to be taken into account. Though experimental demonstration [52-54] already indicated very promising loss level (smaller than 3dB/cm), we consider here a conservative higher waveguide loss level of 8dB/cm. With all the other parameters being fixed (i.e. $p = 150mW$, ring $L = 314\mu m$ and $\gamma = 30m^{-1}W^{-1}$), we used again our model to evaluate the silicon soliton frequency comb properties with the dispersion curves of Figure 7 (a) and 7 (b). The corresponding results are shown in 8 (a)-(c) and 8 (d)-(f), respectively. In Figure 8 (a) and (d), we obtain a new initial comb generation stage, where the primary frequency lines are way farther away from the pump line as compared to figure 4 (a). The following mode-locking multi-soliton stage with cascaded sub combs is shown in Figure 8 (b) and (e), which is much better than the one of figure 4 (b). There is no doubt that the comb bandwidth in stage of the final single-soliton frequency comb in Figure 8 (c) and (f) is similarly broader. Bandwidth improvement is thus clearly witnessed with the introduction of self-adaptive boundary.

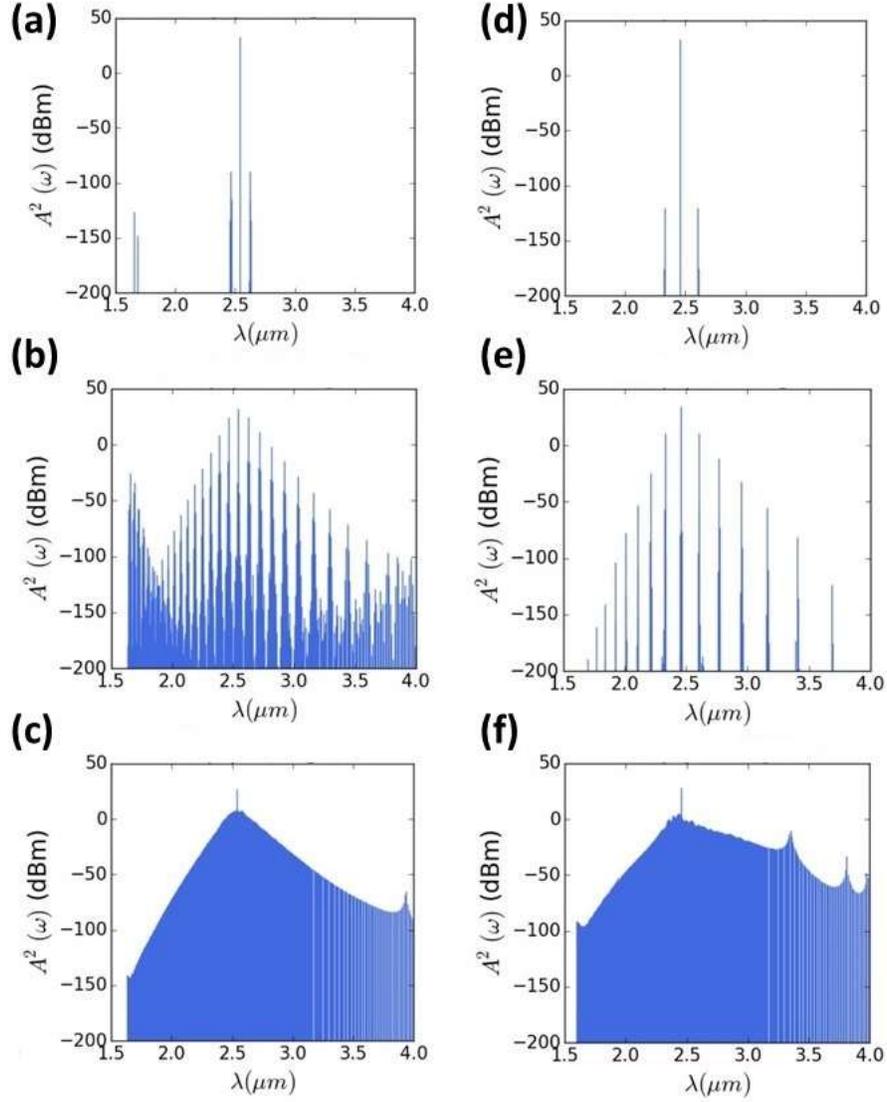

**Fig. 8.** (a) and (d) Snapshot of the generation of primary frequency lines due to the degenerate four-wave mixing, corresponding to dispersion in Figure 7 (a) and 68 (b). (b) and (f) A later snapshot where multi frequency lines by are generated by cascaded four wave mixing, together with mode-locking multi soliton processes. (c) and (f) corresponding single-soliton frequency comb.

**CONCLUSION**

We demonstrate that due to strong dispersion variations, optical frequency comb based on high index contrast waveguides generally display relative limited bandwidths and we propose a new and general approach to trim and expand frequency comb in such high index contrast optical platforms by using self-adaptive boundary waveguides enabled by sub-wavelength index engineering. As a matter of illustration and prime example, this approach enabled us to flatten the dispersion of silicon (on insulator) waveguides and extend the window of needed low-anomalous dispersion, then theoretically to demonstrate silicon frequency combs with improved bandwidth in several different waveguide configurations, e.g. a 3dB bandwidth of 100nm in the $2.5 \mu m$ wavelength. This new strategy targets to address the dispersion issue in high-index-contrast platforms and can be used to explore versatile nonlinear applications in which the manipulation of energy spacing and phase matching is of primary significance. We believe that an application of this new quasi-universal approach will have very broad applications in integrated non-linear photonics for the realization of frequency combs, supercontinuum sources, and other non-linear effects applicable to signal processing, metrology and spectroscopy on a chip.

**SUPPLEMENTARY MATERIAL**

**ACKNOWLEDGMENT**

**REFERENCES**

(1) G. P. Agrawal, Nonlinear Fiber Optics (Academic Press, Boston, 1989).

(2) J. Hansryd, A. Andrekson, M. Westlund, J. Li, P. Hedekvist, "Fiber-based optical parametric amplifiers and their applications," IEEE J. Sel. Top. Quantum Electron. 8, 506 (2002).

(3) 3J. Leuthold, C. Koos and W. Freude, "Nonlinear silicon photonics," Nat. Photon. 4, 535 (2010).

(4) M. Borghi, C. Castellan, S. Signorini, A. Trenti and L. Pavesi, "Nonlinear silicon photonics," J. Opt. 19, 093002 (2017).

(5) M. A. Foster, A. C. Turner, J. E. Sharping, B. S. Schmidt, M. Lipson, A. L. Gaeta, "Broadband optical parametric gain on a silicon photonic chip," Nature 441, 960 (2006).

(6) A. C. Turner-Foster, M. A. Foster, R. Salem, A. L. Gaeta and M. Lipson, "Frequency conversion over two-thirds of an octave in silicon nanowaveguides," Opt. Exp. 18, 1904 (2010).

(7) X. Liu, R. M. O. Jr, Y. A. Vlasov and W. M. J. Green, "Mid-infrared optical parametric amplifier using silicon nanophotonic waveguides," Nat. Photon. 4, 557 (2010).

(8) S. Zlatanovic, J. S. Park, S. Moro, J. M. C. Boggio, I. B. Divliansky, N. Alic, S. Mookherjea and S. Radic, "Mid-infrared wavelength conversion in silicon waveguides using ultracompact telecom-band-derived pump source," Nat. Photon. 4, 561 (2010).

(9) X. Liu, B. Kuyken, G. Roelkens, R. Baets, R. M. O. Jr and W. M. J. Green, "Bridging the mid-infrared-to-telecom gap with silicon nanophotonic spectral translation," Nat. Photon. 6, 667 (2012).

(10) P. Del'Haye, A. Schliesser, O. Arcizet, T. Wilken, R. Holzwarth and T. J. Kippenberg, "Optical frequency comb generation from a monolithic microresonator". Nature 450:1214, 2007.

(11) Y. Okawachi, K. Saha, J. S. Levy, Y. H. Wen, M. Lipson and A. L. Gaeta1, "Octave-spanning frequency comb generation in a silicon nitride chip". Opt. Lett 36: 3398, 2011.

(12) T. Herr, K. Hartinger, J. Riemensberger, C. Y. Wang, E. Gavartin, R. Holzwarth, M. L. Gorodetsky and T. J. Kippenberg, "Universal formation dynamics and noise of Kerr-frequency combs in microresonators". Nat. Photon. 6: 480, 2012.

(13) T. Herr, V. Brasch, J. D. Jost, C. Y. Wang, N. M. Kondratiev, M. L. Gorodetsky and T. J. Kippenberg, "Temporal solitons in optical microresonators". Nat. Photon. 8: 145, 2014.

(14) A. G. Griffith, R. K.W. Lau, J. Cardenas, Y. Okawachi, A. Mohanty, R. Fain, Y. Ho D. Lee, M. Yu, C. T. Phare, C.B. Poitras, A. L. Gaeta and M. Lipson, "Silicon-chip mid-infrared frequency comb generation". Nat. Comm. 6: 6299, 2014.

(15) V. Brasch, M. Geiselmann, T. Herr, G. Lihachev, M. H. P. Pfeiffer, M. L. Gorodetsky and T. J. Kippenberg, "Photonic chip–based optical frequency comb using soliton Cherenkov radiation". Science 351: 357, 2016.

(16) X. Xue, F. Leo, Y. Xuan, J. A. Jaramillo-Villegas, P. Wang, D. E. Leaird, M. Erkintalo, M. Qi, and A. M. Weiner, "Second-harmonic-assisted four-wave mixing in chip-based microresonator frequency comb generation," Light: Science & Applications 6, e16253 (2017).

(17) H. Guo, C. Herkommer, A. Billat, D. Grassani, C. Zhang, M. H. P. Pfeiffer, W. Weng, C. Brès and T. J. Kippenberg, "Mid-infrared frequency comb via coherent dispersive wave generation in silicon nitride nanophotonic waveguides," Nat. Photon. 12, 330 (2018).

(18) N. Singh, M. Xin, D. Vermeulen, K. Shtyrkova, N. Li, P. T. Callahan, E. S. Magden, A. Ruocco, N. Fahrenkopf, C. Baiocco, B. P. Kuo, S. Radic, E. Ippen, F. X. Kärtner, and M. R. Watts, "Octave-spanning coherent supercontinuum generation in silicon on insulator from 1.06 μm to beyond 2.4 μm," Light: Science & Applications 7, 17131 (2018).

(19) B. Stern, X. Ji, Y. Okawachi, A. L. Gaeta and M. Lipson, "Battery-operated integrated frequency comb generator". Nature 562:401, 2018.

(20) T. J. Kippenberg, A. L. Gaeta, M. Lipson, M. L. Gorodetsky, "Dissipative Kerr solitons in optical microresonators," Nat. Photon. 361:567, 2018.

(21) M. A. Foster, A. C. Turner, J. E. Sharping, B. S. Schmidt, M. Lipson, A. L. Gaeta, "Broad-band optical parametric gain on a silicon photonic chip," Nature 441, 960 (2006).

(22) M. A. Foster, A. C. Turner, R. Salem, M. Lipson and A. L. Gaeta, "Broad-band continuous-wave parametric wavelength conversion in silicon nanowaveguides," Opt. Exp. 15, 12949 (2007).

(23) D. W. M. Laughlin, J. V. Moloney, A. C. Newell, "Solitary waves as fixed points of infinite-dimensional maps in an optical bistable ring cavity". Phys. Rev. Lett. 51: 75, 1983.

(24) A. C. Turner-Foster, M. A. Foster, R. Salem, A. L. Gaeta and M. Lipson, "Frequency conversion over two-thirds of an octave in silicon nanowaveguides," Opt. Exp. 18, 1904 (2010).

(25) J. Zhang, C. Alonso-Ramos, L. Vivien, S. He and Eric Cassan, "Self-adaptive waveguide boundary for inter-mode four-wave mixing," *Journal of Selected Topics in Quantum Electronics*, accepted.

(26) M. Eichenfield, J. Chan, R. M. Camacho, K. J. Vahala and O. Painter, "Optomechanical crystals," Nature. 462, 78 (2009).

(27) F. Alpeggiani, L. C. Andreani and D. Gerace, "Effective bichromatic potential for ultra-high Q-factor photonic crystal slab cavities," Appl. Phys. lett. 107, 261110 (2015).

(28) A. Simbula, M. Schatzl, L. Zagaglia, F. Alpeggiani, L. C. Andreani, F. Schäffler, T. Fromherz, M. Galli and D. Gerace, "Realization of high-Q/V photonic crystal cavities defined by an effective Aubry-André-Harper bichromatic potential," Appl. Phys. lett. 2, 056102 (2017).

(29) A. Pasquazi, M. Peccianti, L. Razzari, D. J. Moss, S. Coen, M. Erkintalo, Y. K. Chembo, T. Hansson, S. Wabnitzg, P. Del'Haye, X. Xue, A. M. Weiner, R. Morandotti, "Micro-combs: A novel generation of optical sources," Physics Reports 728:1, 2018.

(30) N. Picqué and T. W. Hänsch, "Frequency comb spectroscopy," Nat. Photon. 13:146, 2019.

(31) P. Feng, J. Kang, S. Tan, Y. Ren, C. Zhang and K. K. Y. Wong, "Dual-comb spectrally encoded confocal microscopy by electro-optic modulators," Opt. lett. 44:2919 (2019).

(32) J. M. Dudley, G. Genty, S. Coen, "Supercontinuum generation in photonic crystal fiber". Reviews of Modern Physics. 78: 1135, 2006.

(33) V. E. Zakharov and A. B. Shabat, "Exact theory of two-dimensional self-focusing and one-dimensional self-modulation of waves in nonlinear". SOVIET PHYSICS JETP 34: 62, 1972.

(34) I. V. Barashenkov and Yu. S. Smirnov, "Existence and stability chart for the ac-driven, damped nonlinear Schrödinger solitons". Phys. Rev. E 54: 5707, 1996.

(35) K. Ikeda, "Multiple-valued stationary state and its instability of the transmitted light by a ring cavity system". Opt. Comm. 30: 257, 1979.

(36) L. Lugiato and R. Lefever, "Spatial Dissipative Structures in Passive Optical Systems". Phys. Rev Lett. 58: 2209, 1987.

(37) L. A. Lugiato, F. Prati, M. L. Gorodetsky and T. J. Kippenberg, "From the Lugiato–Lefever equation to microresonator based soliton Kerr frequency combs". Trans. R. Soc. A 376: 20180113, 2018.

(38) M. Haelterman, S. Trillo and S. Wabnitz, "Dissipative modulation instability in a nonlinear dispersive ring cavity ". Opt. Comm. 91: 401, 1992.

(39) A. B. Matsko, A. A. Savchenkov, W. Liang, V. S. Ilchenko, D. Seidel, and L. Maleki, "Mode-locked Kerr frequency combs". Opt. Comm. 36: 2845, 2011.

(40) S. Coen and M. Erkintalo, "Universal scaling laws of Kerr frequency combs". Opt. Lett. 38: 1790, 2013.

(41) S. Coen, H. G. Randle, T. Sylvestre and M. Erkintalo, "Modeling of octave-spanning Kerr frequency combs using a generalized mean-field Lugiato–Lefever model". Opt. Lett. 38: 37, 2013.

(42) T. Hansson and S. Wabnitz, "Dynamics of microresonator frequency comb generation: models and stability". Nanophotonics 5: 231, 2016.

(43) M. R. E. Lamont, Y. Okawachi, and A. L. Gaeta, "Route to stabilized ultrabroadband microresonator-based frequency combs ". Opt. Exp. 38: 3478, 2013.

(44) H. Guo, M. Karpov, E. Lucas, A. Kordts, M. H. P. Pfeier, V. Brasch, G. Lihachev, V. E. Lobanov, M. L. Gorodetsky and T. J. Kippenberg, "Universal dynamics and deterministic switching of dissipative Kerr solitons in optical microresonators ". Nat. Phys. 13: 94, 2017.

(45) Y. K. Chembo, D.V. Strekalov, and N. Yu, "Spectrum and Dynamics of Optical Frequency Combs Generated with Monolithic Whispering Gallery Mode Resonators". Phys. Rev. Lett. 104: 103902, 2010.

(46) Y. K. Chembo and N. Yu, "Modal expansion approach to optical-frequency-comb generation with monolithic whispering-gallery-mode resonators". Phys. Rev. A 82: 033801, 2010.

(47) R. K. W. Lau, M. R. E. Lamont, Y. Okawachi, and A. L. Gaeta, " Effects of multiphoton absorption on parametric comb generation in silicon microresonators". Opt. Lett. 40: 2778, 2015.

(48) D. Ortega, J. M. Aldariz, J. M. Arnold and J. Stewart Aitchison, "Analysis of "Quasi-Modes" in Periodic Segmented Waveguides," Journal of Light. Tech. 17, 369 (1999).

(49) R. Halir, P. J. Bock, P. Cheben, A. Ortega-Moñux, C. Alonso-Ramos, J. H. Schmid, J. Lapointe, D. Xu, J. G. Wanguemert-Perez, I. Molina-Fernández and S. Janz, "Waveguide sub-wavelength structures: a review of principles and applications," Laser Photonics Rev. 9,25 (2015).

(50) P. Cheben, R. Halir, J. H. Schmid, H. A. Atwater and D. R. Smith, "Subwavelength integrated photonics," Nature. 560, 565 (2018).

(51) J. S. Penadés, C. Alonso-Ramos, A. Z. Khokhar, M. Nedeljkovic, L. A. Boodhoo, A. Ortega-Moñux, I. Molina-Fernández, P. Cheben, and G. Z. Mashanovich, "Suspended SOI waveguide with sub-wavelength grating cladding for mid-infrared," Opt. Lett. 39, 5661 (2014).

(52) J. S. Penades, A. Ortega-moñux, M. Nedeljkovic, J. G. Wangüemert-pérez, R. Halir, A. Z. Khokhar, C. Alonso-ramos, Z. Qu, I. Molina-fernández, P. Cheben and G. Z. Mashanovich, "Suspended silicon mid-infrared waveguide devices with subwavelength grating metamaterial cladding". Opt. Exp. 24: 022908, 2016.

(53) H. Lin, Z. Luo, T. Gu, L. C. Kimerling, K. Wada, A. Agarwal and J. Hu, "Mid-infrared integrated photonics on silicon: a perspective". Nanophotonics 7: 393, 2018.

(54) S. A. Miller, M. Yu, X. ji, A. G. griffith, J. Cardenas, A. Gaeta and M. Lipson, "Low-loss silicon platform for broadband mid-infrared photonics". Optica 4: 707, 2017.